\documentclass[]{spie}  

\usepackage{multirow}
\usepackage{framed,multirow} 
\usepackage{amsmath,amsfonts,amssymb} 
\usepackage{wrapfig}  
\usepackage{graphicx}
\usepackage[colorlinks=true, allcolors=blue]{hyperref}

\usepackage{array, makecell}
\usepackage{boldline}
\usepackage{colortbl}
\usepackage{color}
\usepackage{xcolor}
\usepackage[final]{microtype}

\begin{document}
\title{Correlation Ratio for Unsupervised Learning of Multi-modal Deformable Registration}

%
%


\author[a]{Xiaojian Chen}
\author[b]{Yihao Liu}
\author[a]{Shuwen Wei}
\author[a]{\\Aaron Carass}
\author[a]{Yong Du}
\author[a]{Junyu Chen}
\affil[a]{Johns Hopkins University, Baltimore, MD 21218, USA}
\affil[b]{Vanderbilt University, Nashville, TN 37235, USA}

\authorinfo{Further author information: (Send correspondence to Junyu Chen: jchen245@jhmi.edu)}

 


%
\maketitle              
\begin{abstract}
In recent years, unsupervised learning for deformable image registration has been a major research focus. 
This approach involves training a registration network using pairs of moving and fixed images, along with a loss function that combines an image similarity measure and deformation regularization. 
For multi-modal image registration tasks, the correlation ratio has been a widely-used image similarity measure historically, yet it has been underexplored in current deep learning methods.
Here, we propose a differentiable correlation ratio to use as a loss function for learning-based multi-modal deformable image registration. 
This approach extends the traditionally non-differentiable implementation of the correlation ratio by using the Parzen windowing approximation, enabling backpropagation with deep neural networks.
We validated the proposed correlation ratio on a multi-modal neuroimaging dataset.
In addition, we established a Bayesian training framework to study how the trade-off between the deformation regularizer and similarity measures, including mutual information and our proposed correlation ratio, affects the registration performance.
The source code is freely available at: \url{bit.ly/3XTJrJh}.





\keywords{Deformable Image Registration, Multi-modality, Unsupervised registration, Correlation Ratio}
\end{abstract}

\section{Introduction}
Deformable image registration~(DIR) is an active area of research in medical image analysis focused on aligning corresponding anatomical or functional structures in different images under non-rigid deformations~\cite{sotiras2013deformable, rigaud2019deformable, chen2023survey}.
As a critical application of DIR, multi-modal medical image registration plays a crucial role in clinical diagnosis and treatment planning, enabling the integration of information acquired by different imaging devices or protocols~\cite{chen2017mia, jiang2021review, velesaca2024multimodal}.
Since ground truth deformations are rarely available for supervised learning, deformable registration is traditionally formulated as an optimization task, where the energy function to be optimized involves an intensity similarity term and a deformation regularizer term.
In the context of multi-modal registration, commonly used similarity measure includes mutual information~(MI)~\cite{wells1996mia, viola1997alignment, meyer1997demonstration}, modality independent neighbourhood descriptor~(MIND)~\cite{heinrich2012mind}, and correlation ratio~(CR)~\cite{roche1998correlation}. 
Additionally, a transformation regularizer (e.g., bending energy~\cite{rueckert1999nonrigid}) is used to enforce deformation smoothness, thereby promoting realistic and plausible deformations. 

Over the past decade, advancements in deep learning have shown considerable promise in various medical image analysis tasks, including image registration. These methods not only improve registration accuracy over traditional optimization-based approaches but also increase efficiency by alleviating the need for iterative optimization processes.
More recently, unsupervised deep learning methods have been applied to multi-modal image registration, using either MIND or MI as the similarity measure in the loss function.~\cite{guo2019multi, qiu2021learning, hansen2021graphregnet, mok2021conditional, xu2020adversarial, blendowski2021weakly, liu2022mmmi, liu2024jmi}.
However, to the best of our knowledge, the correlation ratio~(CR), though widely used as a multi-modal similarity measure in traditional image registration methods, has not been used in deep learning-based medical image registration.
This omission likely stems from the fact that the original implementation of CR
involves discretely iterating through intensity ranges, making it non-differentiable and unsuitable for direct use in deep learning-based methods.
Additionally, the trade-off between the CR and deformation regularization remains unclear and requires further study for successful practical usage.
Previous works have either arbitrarily set the weighting hyperparameter $\lambda$ for the deformation regularizer~\cite{balakrishnan2019voxelmorph, kim2021cyclemorph, chen2022transmorph} or determined it through a grid search~\cite{ciardo2013role, kang2019optimized}.
However, none of these studies have fully explored how $\lambda$ impacts the registration results.

In this work, we introduce a differentiable CR for learning-based multi-modal deformable registration.
Our work is inspired by recent a Parzen-window-based MI method used in learning-based deformable image registration~\cite{guo2019multi}.
We compared our differentiable CR and MI using the multi-modal neuroimaging IXI dataset. 
In addition, we established a Bayesian training framework to study the optimal trade-off between the deformation regularizer and similarity measures, including MI and our differentiable CR.
To benefit further research, we make our source code publicly available. 
\begin{figure}[!tb]
\centering
    \includegraphics[width=0.85\columnwidth]{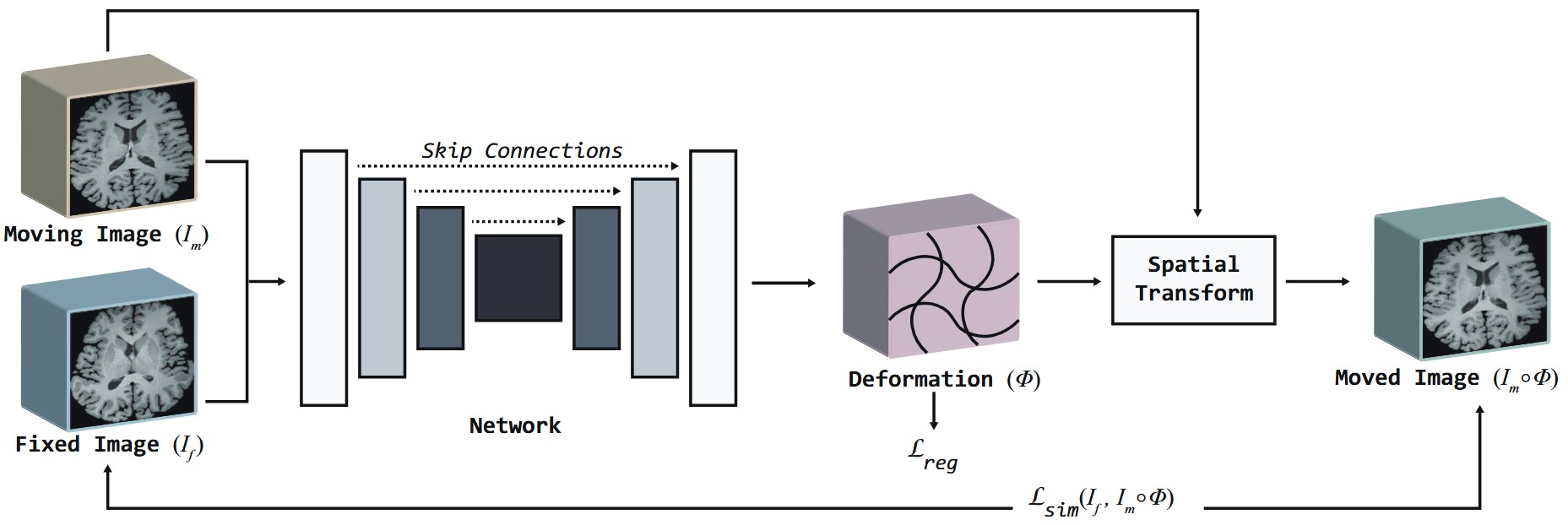}
    \caption{The general framework of our model.} 
    \label{fig:framework}
\end{figure}

\section{Methodology}

\subsection{Preliminaries}
\label{sec:preliminaries}
Traditionally, CR is calculated using a discrete approach by iterating over intensity values, which is non-differentiable and unsuitable for deep neural network (DNN) training. To overcome this limitation, a continuous approach using Parzen window density estimation can be employed to enable auto-differentiation in DNNs. 
The Parzen window density estimation~\cite{kwak2002input} has been used in the differentiable implementation of MI~\cite{guo2019multi,qiu2021learning} to approximate the probability density function by placing Gaussian kernels around the centers of the intensity bins:
\begin{equation}
\label{eqn:parzen}
    \omega_{ik}(X;h,\text{bin}) = \frac{1}{h \sqrt{2\pi}} \exp \left( -\frac{(x_i - \text{bin}_k)^2}{2h^2} \right),
\end{equation} 
where \(\omega_{ik}\) denotes the contribution to the $k$-th intensity bin for the $i$-th voxel $x_i$ in image $X$, and \(h\) denotes the standard deviation of the Gaussian kernel.
We leverage this approximation to derive a differentiable loss function that maximizes CR for multi-modal registration.

\subsection{Differentiable Correlation Ratio in DL registration}
\label{sec:cr}
Let $I_{m}$ and $I_f$ be the moving and fixed volumes, respectively, defined on a 3-D spatial domain $\Omega\subset\mathbb{R}^3$.
Although $I_{m}$ and $I_f$ can belong to either the same or different modalities, our focus here is on the latter scenario. The moving image \(I_m\) is warped using a deformation field $\phi$, generated by a registration DNN.
The DNN is trained using CR as the similarity measure between modalities to ensure that the transformation $\phi$ aligns $I_{m} \circ \phi$ closely with the fixed image $I_f$.

Given two images $X$ and $Y$, the CR between these images is defined as:
\begin{equation}
    \eta(Y|X) = \frac{Var(\mathbb{E}(Y|X))}{Var(X)},
\end{equation}
where $\mathbb{E}(Y|X)$ is the conditional expectation of $Y$ given $X$. 
For a specific intensity bin $k$, this conditional expectation can be approximated by a weighted mean $\Bar{y}_k = \frac{\sum_{i}\omega_{ik}y_i}{\sum_{i}\omega_{ik}}$, where $y_i$ denotes the $i$-th voxel in $Y$, and $\omega_{ik}$ is the weighting function defined in Eqn.~\ref{eqn:parzen}, calculated from image $X$.
The variance of this conditional expectation is calculated as $Var(\mathbb{E}(Y|X)) = \sum_kn_k(\Bar{y}_k - \Bar{y})^2$, where $n_k=\frac{\sum_iw_{ik}}{\sum_i\sum_kw_{ik}}$ represents the normalized weight reflecting the proportion of the total weighting function contributions within the $k$-th intensity bin.
Additionally, $\Bar{y}=\frac{1}{|\Omega|}\sum_iy_i$ denotes the mean intensity of image $Y$. Finally, $Var(X)$ is the variance of the intensity values of $X$ given by $Var(X)=\frac{1}{|\Omega|}\sum_i(x_i-\Bar{x})^2$, where $\Bar{x}$ is the mean intensity of image $X$.

With this formulation, we construct the differentiable CR for the deformed moving and fixed images (i.e., $I_m\circ\phi$ and $I_f$) as follows: 
\begin{equation}
    \mathcal{L}_{CR}(I_f, I_{m} \circ \phi) = -\frac{1}{2} \left( \eta(I_f|I_{m} \circ \phi) + \eta(I_{m} \circ \phi|I_f) \right).
    \label{eq:cr}
\end{equation}
Note that we opt for a symmetric loss function for training due to the inherent asymmetry of CR (i.e., $\eta(I_f|I_{m} \circ \phi) \neq \eta(I_{m} \circ \phi|I_f)$).

\subsection{Registration framework}
\label{sec:train_loss}
Our deformable registration framework follows the conventional pipeline for unsupervised learning-based registration~\cite{balakrishnan2019voxelmorph,chen2022transmorph}.
Our proposed differentiable CR is agnostic to the network architecture.
For demonstration, two DNN backbones, TransMorph~\cite{chen2022transmorph, chen2022unsupervised} and VoxelMorph~\cite{balakrishnan2019voxelmorph}, were investigated in this paper. 
The overall loss function to train the DNN is defined as follows:
\begin{equation}
    \mathcal{L}(I_f, I_{m} \circ \phi) = \mathcal{L}_{\textit{CR}}(I_f, I_{m} \circ \phi) + \lambda \mathcal{L}_{\textit{Reg}}(\phi).
    \label{eq:CtRL}
\end{equation}
where $\mathcal{L}_{\textit{Reg}}$ is the diffusion regularization placed on the deformation field $\phi$ that encourages the displacement value in a location to be similar to the values in its neighboring locations~\cite{balakrishnan2019voxelmorph}.
It aims to enforce smoothness in the deformation field, and is formulated as $\mathcal{L}_{Reg}(\phi) = \frac{1}{|\Omega|} \sum_{p \in \Omega} \|\nabla \mathbf{u}(p)\|^2$, where $\nabla \mathbf{u}(p)$ represents the spatial gradients of each component of the displacement field $\mathbf{u}$ at voxel location $p$. The parameter $\lambda$ is the weighting hyperparameter for $\mathcal{L}_{Reg}$, and its optimal value must be determined through a thorough search. Details on exploring this trade-off for CR are discussed in the Section~\ref{sec:Opt}.


\begin{figure}[!tb]
\centering
    \includegraphics[width=0.8\columnwidth]{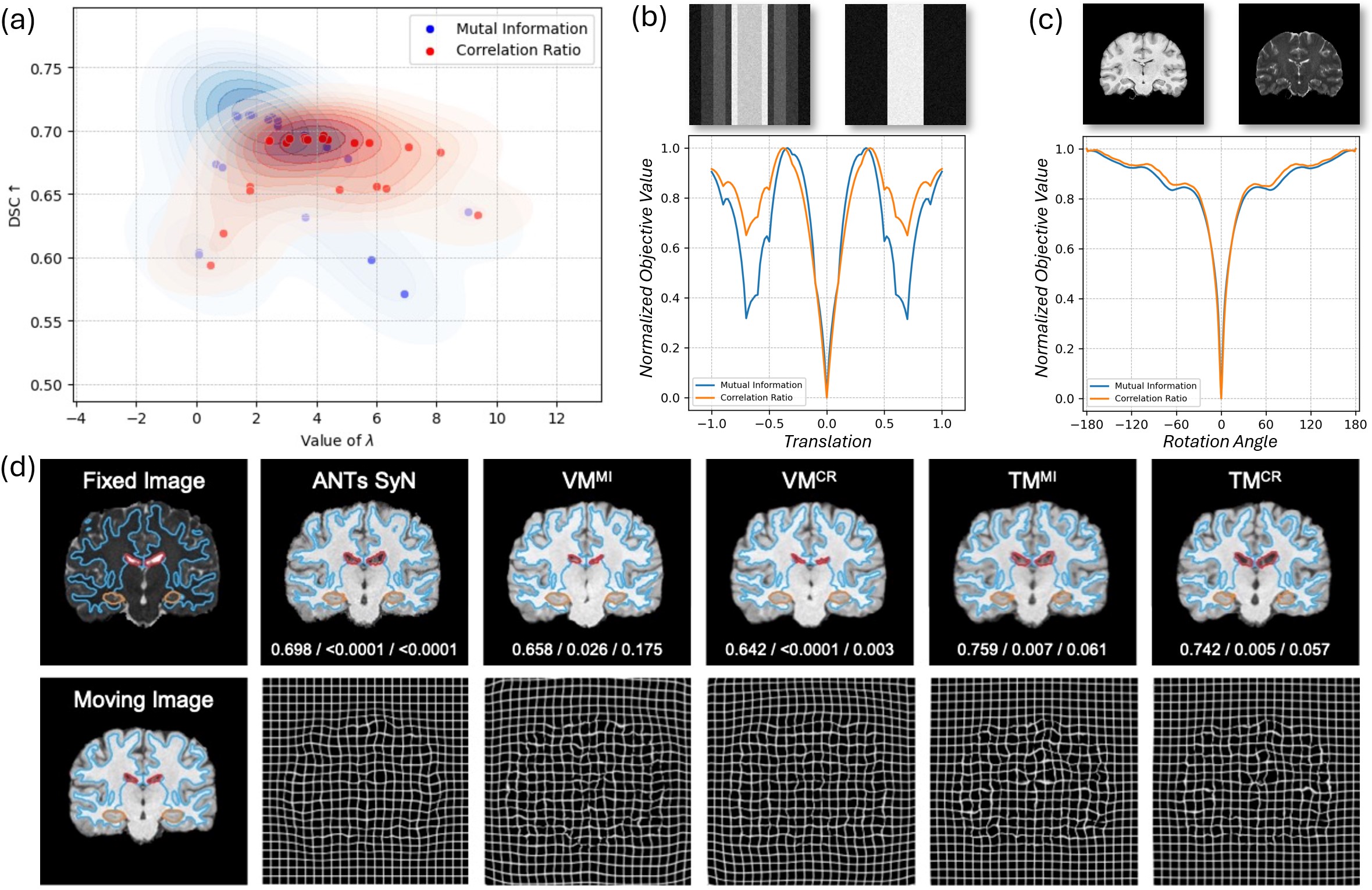}
    \caption{Quantitative and qualitative results comparing different deformable registration methods. \textbf{(a)}~Hyperparameter optimization results for $\lambda$. \textbf{(b)}~and \textbf{(c)}~Objective value versus translation (b) and rotation (c) between the images shown on top. \textbf{(d)}~Example MR coronal slices extracted from input pairs ($I_f$, $I_m$), and the resulting ($I_{m} \circ \phi$) for different methods with optimal $\lambda$.} 
    \label{fig:result_opt}
\end{figure}

\section{Results}

\subsection{Experimental Setup}
\label{sec:Exp}
\noindent\textbf{Data and Preprocessing.} 
We evaluated the proposed method for multi-modal T1w-to-T2w image registration using the IXI dataset~\cite{ixi}, which consists of 575 subjects. The dataset was divided into 400 subjects for training, 40 for validation, and 135 for testing.
Each subject has a pair of T1w and T2w scans, cropped to $160\times192\times224$.
During training, subjects were randomly selected from the training set to serve as moving and fixed images, with the two scans in each pair coming from different subjects. All MRI scans underwent N4 inhomogeneity correction~\cite{tustison2010n4itk} and were pre-aligned with an atlas image using a rigid transformation. Intensity values for each scan were normalized through white matter peak normalization~\cite{reinhold2019evaluating}. 
For evaluating the registration performance, we generated whole brain segmentation of 133 anatomical structures using SLANT~\cite{huo20193d}.

\noindent\textbf{Baselines and Implementation Details.}
We compared our differentiable CR to MI~\cite{guo2019multi} across learning-based registration methods including TransMorph~\cite{chen2022transmorph} and VoxelMorph~\cite{balakrishnan2019voxelmorph}. Additionally, we included the results of an optimization-based method, SyN~\cite{avants2008symmetric}, using MI.
The performance is evaluated in terms of both \textit{accuracy} and the transformation \textit{regularity}. \textit{Accuracy} is assessed by Dice similarity coefficient (DSC)~\cite{dice1945measures}, calculating the overlap between the anatomical segmentation of the fixed image and the segmentation transformed by \( \phi \) in the moving image. The transformation \textit{regularity} is evaluated by non-diffeomorphic volume (NDV)~\cite{liu2024finite} and the ratio of voxels with the negative determinant of the Jacobian ($|J(\phi)|\leq0$). 
For each dataset, the hyperparameter for deformation regularization (i.e., $\lambda$) was carefully tuned using Optuna~\cite{akiba2019optuna} for Bayesian optimization of the Tree-Structured Parzen Estimator algorithm.
We conducted 20 trials for TransMorph and VoxelMorph to search for the best $\lambda$ for both MI and CR, with each trial running for 300 epochs.

\subsection{Hyperparameter Optimization} 
\label{sec:Opt}
Figure~\ref{fig:result_opt}~(a) presents a scatter plot of the mean DSC scores versus $\lambda$ resulting from the Bayesian hyperparameter optimization, separately for MI and CR on TransMorph.
Note that some low DSC values for specific $\lambda$ values were due to pruning by Optuna~\cite{akiba2019optuna}.
To provide an intuitive comparison of MI and CR performance, we also include contour plots composed of scatter points. It is evident that when using MI, the optimal DSC values fall within a narrower range of $\lambda$ values, indicating high sensitivity and a need for precise selection of $\lambda$. In contrast, when using CR, the optimal $\lambda$ results in a slightly lower DSC score compared to that of MI, but it retains more stable DSC values across a broader range of $\lambda$ values. It is noteworthy that the optimal $\lambda$ values differ between VoxelMorph (4.5 for MI and 7.7 for CR) and TransMorph (1.7 for MI and 4.2 for CR), suggesting that different DNNs may favor different $\lambda$. This leads to an area of interest that requires further research.

\begin{table}[!tb]
\centering
 \fontsize{9}{11}\selectfont
\caption{Quantitative results on the inter-modal brain registration task. We report averaged DSC, the percentage of NDV~\cite{liu2024finite}, and the percentage of voxels with $|J(\phi)| \leq 0$, with the corresponding standard deviations.}
\begin{tabular}{c|c|c|c|c}
    \hline
    \textbf{Method} & \textbf{$\lambda$} & \textbf{DSC} $\uparrow$ & \textbf{\% NDV} $\downarrow$ & \textbf{\% of $|J(\phi)| \leq 0$} $\downarrow$ \\ \hline
    \rowcolor{green!20!white}
     {ANTs SyN (MI)}  & - & $0.636 \pm 0.048$ & $< 0.0001$ & $< 0.0001$\\ 
     \hline
     {VoxelMorph (MI)} & 4.5 & $0.594 \pm 0.071$ & $0.004 \pm 0.004$ & $0.063 \pm 0.051$\\
     \rowcolor{green!20!white}
     {TransMorph (MI)} & 1.7 & $0.713 \pm 0.028$ & $0.009 \pm 0.005$ & $0.108 \pm 0.063$ \\
    \hline
     VoxelMorph (\textit{CR}) & 7.7 & $0.594 \pm 0.063$ & $0.002 \pm 0.004$ & $0.006 \pm 0.009$ \\
     \rowcolor{green!20!white}
     TransMorph (\textit{CR}) & 4.2 & $0.691 \pm 0.031$ & $0.014 \pm 0.009$ & $0.119 \pm 0.056$ \\ 
     \hline
    
\end{tabular}
    \label{tab:our_model_table_reply}
\end{table}

\subsection{Qualitative and Quantitative Results}
Qualitative results are shown in Fig.~\ref{fig:result_opt}~(d). The first row displays the deformed results from different methods, with three anatomical regions contoured: the cerebral white matter in blue, the lateral ventricles in red, and the hippocampus in yellow.
The scores at the bottom of the images correspond to the DSC, the percentage of NDV, and the percentage of voxels with $|J(\phi)| \leq 0$.
The second row depicts the respective deformation fields $\phi$.
Quantitative results on the test set are shown in Table~\ref{tab:our_model_table_reply}.
When trained with CR, the models achieved slightly lower DSC scores using TransMorph but comparable DSC scores with VoxelMorph compared to those trained with MI. This finding suggests that CR and MI are similarly effective as similarity measures for brain image registration. However, there are merits in adopting CR, as demonstrated by two illustrative examples in Figs.~\ref{fig:result_opt}~(b) and~(c).
These figures show the objective value (the lower, the better) versus the degree of translation and rotation, where CR demonstrated a smoother landscape, making it less likely to get stuck in local minima. Additionally, we compared the computational speed of the two similarity measures on an H100 GPU for an image pair of size $160\times192\times224$, averaged over 100 runs.
CR achieved a computation speed of \textbf{0.026} seconds, while MI achieved \textbf{0.149} seconds, making CR almost \textbf{6-fold} faster than MI.
This significant speed advantage makes CR potentially suitable for applications such as instance-specific optimization and registration guided surgical applications.


\section{Discussion and Conclusion}

In this paper, we proposed a differentiable implementation of CR for multi-modal deformable image registration, utilizing the Parzen windowing approximation to avoid discretely counting intensity bins, inspired by the recently developed Parzen-window-based MI~\cite{guo2019multi}.
The hyperparameter trade-off between the similarity measure and the deformation regularizer, previously subject to arbitrary choice, underwent extensive experimentation to determine the optimal value.
The proposed method was validated on T1w-to-T2w registration using the IXI dataset.
Experimental results demonstrate that the method is applicable to different networks, performing competitively with MI but computing significantly faster. Future work will focus on comparing additional similarity measures for multi-modal registration.


%
%
%

\acknowledgments 
This study was supported by the grants from the National Institutes of Health (NIH), United States. The work was made possible in part by the Johns Hopkins University Discovery Grant (Co-PI: J. Chen, Co-PI: A. Carass).


%
%
%
\bibliography{miccai24}

\bibliographystyle{spiebib} 

\end{document}